\documentstyle[12pt,aasms4]{article}

\def\lta{{\>\rlap{\raise2pt\hbox{$<$}}\lower3pt\hbox{$\sim$}\>}}
\def\gta{{\>\rlap{\raise2pt\hbox{$>$}}\lower3pt\hbox{$\sim$}\>}}

\lefthead{Stiavelli et al.\ }
\righthead{The Interacting Galaxy Pair NGC~454}

\begin{document}
 
\title{WFPC2 Observations of NGC~454: an Interacting Pair of Galaxies
\footnote{Based on observations with the NASA/ESA Hubble Space
Telescope, obtained at the Space Telescope Science Institute, which is
operated by Association of Universities for Research in Astronomy,
Inc.\ (AURA), under NASA contract NAS5-26555}}

\author{M.~Stiavelli\footnote{On leave from the Scuola Normale
Superiore, Piazza dei Cavalieri 7, I-56126 Pisa, Italy}$^,$\footnote{On
assignment from the Space Science Dept. of the European Space Agency},
N.~Panagia$^3$}

\affil{Space Telescope Science Institute, 3700 San Martin Drive,
Baltimore, MD 21218} 

\author{C.~Marcella Carollo\footnote{Hubble Fellow} }

\affil{Department of Physics \& Astronomy, The Johns Hopkins
University, Baltimore, MD 21218}

\author{M. Romaniello\footnote{also Scuola Normale Superiore, Pisa},
I. Heyer and S. Gonzaga}

\affil{Space Telescope Science Institute, 3700 San Martin Drive,
Baltimore, MD 21218}

\begin{abstract}
We present WFPC2 images in the F450W, F606W and F814W filters of the
interacting pair of galaxies NGC~454. Our data indicate that the
system is in the early stages of interaction. A population of young
star-clusters has formed around the late component, and 
substantial amounts of gas have sunk into the center of the
earlier component, where it has not yet produced significant visible star
formation or nuclear activity.  We have photometric evidence that the
star-clusters have strong line emission, which indicate the 
presence of a substantial component of hot, massive stars which formed 
less than 5-10$Myrs$ ago.

\end{abstract}
{\it subject headings}: galaxies: elliptical and lenticular, cD ---
galaxies: individual (NGC 454) --- galaxies: star clusters ---
galaxies: interactions

\section{Introduction}

Extensive evidence both from the nearby universe and high redshift
studies shows that galaxies interact and that 
interactions can strongly affect the galactic properties and even alter
the morphological type. Numerical multi-component simulations (Barnes and
Hernquist 1996) have shown that the
interstellar gas rapidly sinks into the cores of the
interacting galaxies, where it cools and forms 
stars. Because of their origin, stars formed in this way are not
expected to have an angular momentum vector aligned with the
main galactic body. Indeed, a significant fraction of
elliptical galaxies possesses kinematically decoupled cores (hereafter
KDC). However, systematic observations of KDC galaxies with the Hubble
Space Telescope (hereafter HST) have shown only indirect evidence for disks
and little evidence for young stellar populations in the nuclear regions
(Carollo et al. 1997a,b).
In addition, the nuclear properties of
KDC galaxies are essentially indistinguishable from
those of galaxies without (known) KDC nuclei. One possibility is that the
phenomena which lead to a KDC actually occur in all galaxies and that
they shape their nuclear properties (Carollo et al. 1997b). 

To investigate these issues we have studied the interacting system NGC 454
(Arp and Madore 1987). Johansson (1988) describes
NGC~454 as a pair consisting of a red elliptical galaxy (Eastern
component, E) and a blue irregular or possibly a disk galaxy (Western
component, W). To the south of W there are three very blue knots (SW,
SE, and S) that were identified by Johansson (1988) as possible
young globular clusters. The evidence for the interacting
nature of NGC~454 comes from the distorted morphology of both
the E and W components, from the fact that both objects
are embedded in a common, low surface brightness halo, and from
spectroscopic and photometric evidence for a young stellar
population in the system. 

We observed NGC~454 with the Wide Field and Planetary Camera 2 (hereafter
WFPC2) aboard HST.   The observations are described in Section 2.
Section 3 is devoted to a description of our data analysis techniques,
while our results are discussed in Section 4.

\section{Observations}

NGC~454 was observed in fine lock during a single orbit on March 6th,
1997. Together with 8 narrow band images of the planetary nebula NGC
2346, these were the first science exposures obtained with WFPC2 after
Servicing Mission 2 (hereafter SM2). We obtained, at gain 7, 2$\times $
500 seconds F450W exposures, 2 $\times $ 140 seconds F606W exposures,
and a 200 plus a  400 seconds F814W exposures. 
Each image was processed with the
standard HST WFPC2 pipeline for bias and dark current
subtraction, and flat fielding. The flat fields used were those
obtained before SM2.  We verified that no change could be seen in
internal flat fields taken before and after SM2 and that no trend was
visible in the sky far from the targets. The two frames available in
each filter were combined using the STSDAS IRAF task CRREJ. 
The central pixel in the elliptical
galaxy in the F606W filter, and the innermost 3 and 5 pixels, respectively, in
the 200 and 400 seconds F814W exposures, saturated the A/D conversion. 

Photometric calibration was carried out in two different ways. 
When studying young stellar populations
around NGC~454 West or areas potentially affected by strong line
emission, we considered directly the HST magnitudes, derived applying
Holtzman et al. (1995) zero-points to the instrumental magnitudes, and
compared them with model calculations in the same WFPC2 magnitude
system (see Section 3.2). Instead, when studying the older population
of NGC~454 East we converted the magnitudes from the WFPC2 system into
the BVI Johnson-Cousins system according to the Holtzman et al. (1995)
synthetic calibration. In fact for young stellar systems, the difference in 
bandpass between the F450W and the F606W filters and the corresponding Johnson
filters, and the effects of dust and emission lines, can make a conversion from
the WFPC2 system to the BVI Johnson-Cousins system very inaccurate.
The use of the pre-SM2 zero-points and
calibration for our post-SM2 data is justified by the throughput
stability demonstrated during WFPC2 SM2 activities. 

In Figure 1a (Plate XA) we display a combined F450W+ F606W+ F814W
image showing the extended common halo and the distorted morphology of
both components. In Figure 1b (Plate XB) we show the F450W-F814W color
map. Visible are the knots of star formation on the W component as well as
the dust in the nucleus of the E component. There are a few blue
pixels in the nucleus of E which could be due to recently formed
stars.

\section{Data Analysis}

In the following we report on:
{\it i)} light and color profiles for E, {\it ii)} point
source photometry for W, and {\it iii)} pixel-by-pixel color-color
diagram analysis for various selected areas (shown in Figure 1).

\subsection{Light and Color Profiles}

Light and color profiles of the E component were derived by using two
independent isophote fitting programs both running within IRAF:
GALPHOT (Franx et al. 1989) and ELLIPSE (in STSDAS). 
Both programs gave essentially the same results. The agreement
for the F814W profile is excellent for all radii larger than 0.14
arcsec. Inside this radius the profile was affected by the saturated
pixels.  For the F450W and F606W profiles there were departures within
about 2 pixels of the center. In particular, such disagreement seemed
to be more serious for the F606W image. The final profiles were
obtained by averaging results from the two codes and are shown in
Figure 2, where we plot the individual light profiles (upper
left panel) and the F450W-F606W and the F606W-F814W color profiles
(bottom panels). The upper right panel contains a color-color plot
(B-V vs V-I); for this latter plot the colors were converted to the
Johnson-Cousins system so that we could compare them with the
theoretical models by Worthey (1994, shown as solid diamonds connected
by a solid line). In the figure we also show the reddening line
(dotted) and an arrow indicating the direction of contamination by
emission lines (W$_{eq}($ H$\alpha)$ = 500\AA\ ). 
The points on the upper left of
the theoretical models are not an artifact of the isophotal fits, since they
are confirmed by inspection of the pixel values in
the nuclear region.
They are probably due to contamination by emission lines.

The theoretical models belong to an age-metallicity sequence from
Worthey (1994). The population models we have considered are: {\it i)}
a 17 Gyrs, [Fe/H]=0.5 population; {\it ii)} a 12 Gyrs, [Fe/H]=0.25
population; {\it iii)} a 8 Gyrs, [Fe/H]=0 model, and {\it iv)} a 5
Gyrs, [Fe/H]=-0.22 model. 
Models at either constant age or constant metallicity fail to reproduce
the observed points, since they produce too small a variation in $B-V$
for a given change in $V-I$. It is only by making the reddest point
both older and more metal rich that we can fit the observations.
In any case, the youngest, least metal rich models
remain redder than the data (particularly in $B-V$). Johansson
(1988), on the basis of the global spectral energy distribution of 
NGC 454-E, claimed that it contained a small fraction of very
young stars. 
An alternative is that NGC 454-E is actually a lenticular galaxy, as suggested
by its global colors and the spectrum published by Johansson (1988).

The light profiles of NGC 454-E are not well fitted by an $R^{1/4}$
law alone, or by an exponential alone. An $R^{1/4}$ law can fit well the outer
parts (with $R_e \simeq 17''$) but falls below the data
inside $\sim 2.5''$.  The profiles can be well fitted by  the sum of an
$R^{1/4}$ plus an exponential law. Given the colors of the object it is
likely that this decomposition reflects the existence of two
physically distinct components: an $R^{1/4}$ bulge and an exponential
disk. The resulting parameters for the (least affected by dust
absorption) $V$ and $I$ filters are: $r_{eV} = 3.1''$,
$V_{F606W,bulge} =13.3$, $r_{eI}=2.4''$, $I_{F814W,bulge} = 12.7$. 
The disk component has an exponential scale
of $10\pm1''$ similar in the two filters.  We have also performed a
Nuker's law fit (Lauer et al. 1995) of the innermost parts of the
F450W and the F606W profiles (the F814W was excluded
because of  the saturated pixels). We find that the average inner slope is
$\gamma = 0.82\pm0.05$ and the average break radius is $r_b = 0.70\pm0.02$.
At the distance of 56 Mpc ($H_0 = 65 $km s$^{-1}$ Mpc$^{-1}$),
the
bulge absolute magnitude is $M_V=-20.4$. For such an object a nuclear
slope of 0.82 is not unusual (Byun et al. 1995).

\subsection{Point Source Photometry}

The point sources near the two galaxies were identified on the WF2 and
WF3 chips in the F814W image (the deepest one) using the DAOFIND
package within IRAF. The search was done with a threshold of $5\sigma$
above the local background.  Subsequently, we carried out
aperture photometry on these sources, retaining  in the final list
only those with a signal-to-noise ratio of at least 3 in all
passbands. 

At the distance of NGC~454 the effective resolution in
the WF chips, $0^{\prime\prime}.15$, corresponds to a linear size of
41~pc. Therefore, it is most likely that all point sources in our
frames be clusters rather than individual stars. This conclusion is
confirmed by their bright magnitudes, i.e. $M_V<-9$. 

In order to derive the properties of these objects, we computed
the time evolution of a single burst of star formation. For this
purpose, we used the FRANEC stellar evolutionary tracks (Brocato
\& Castellani 1993 and Cassisi, Castellani \& Straniero 1994) together
with the Kurucz (1993) model atmospheres. The model colors were
computed in the WFPC2 passbands, using the most up-to-date
instrumental response curves available (SYNPHOT package in STSDAS),
thus, enabling us to make a direct comparison between data and theory.

In Figure 3 we plot a color-color diagram for all point sources (a
total of 233, of which 56 in the WF2  chip and 177 in the WF3 chip)
and compare them to the theoretical evolutionary tracks (heavy solid
line, 3 Myrs to 5 Gyrs).  We also plot for reference the distribution
of colors (i.e. the colors of individual pixels) in galaxies E and W.
It is apparent that while E and W essentially lie on the theoretical
tracks, all point sources are significantly displaced from them. The
effects of reddening and contamination by emission lines were modelled
adopting a screen of dust (with the same properties as in the Milky
Way) and using Johansson's spectra as templates for emission lines'
intensities (Romaniello et al. 1997). The
corresponding vectors are indicated with arrows in Figure 3. It is
clear that both effects are at work to produce the observed point
source displacements.  On average, internal reddenings of about
$E(B-V)\simeq0.4$ and H$\alpha$ equivalent widths in excess of
$W_{eq}(H\alpha)\simeq1000$ \AA\ are required to reproduce the observed
colors. The reason why both values are higher than found by Johansson
(1988) in his study of the compact objects is that here we resolve the
star forming regions much more in detail and, therefore, the
contamination of light from older background populations is much less
severe.  The strong line emission in the spectra of theses clusters
imply a very young age for the ionizing stars, namely less than 10
$Myrs$ and, possibly younger than 5$Myrs$.  These estimates
corresponds to the lifetime of O type stars which are the most
important sources of ionizing radiation among young stars (e.g.
Panagia 1973) and have lifetimes shorter than 10 $Myrs$ (e.g. Iben
1967).

Even before reddening correction, the F606W magnitudes of these
objects are quite bright, ranging from -9.5 up to -14.5 with a median
value of -11.3. The luminosity function (see Figure 3) resembles those
found for  star-clusters in interacting galaxies in more advanced
stages of interaction (Whitmore and Schweizer 1995, Whitmore et al.
1993). The corresponding masses, estimated under the conservative
assumption that they consist of young population only, are around
$1.8\times 10^5~M_\odot$ with a dispersion of a factor of 3 either
direction.  The total mass in compact clusters turns out to be
somewhat higher than $5\times 10^7~M_\odot$. This is a little less than
1\% of the entire mass of NGC~454 West ($\sim10^{10}~M_\odot$ as
estimated from its total visual magnitude, -18.8, and adopting a mass
to luminosity ratio of 3). On the other hand, such mass may represent
a sizable fraction of the entire gas content of NGC~454 West (say,
$M_{gas}\sim0.1\times M_{total}$), confirming that star formation
induced by galaxy interactions can be very efficient.  A detailed 
discussion of these aspects will be presented in a forthcoming paper
(Romaniello et al. 1997).

\subsection{Individual Pixel Color Analysis}

The properties of the stellar populations in the two galaxies and
their uniformity were studied considering the color-color plots of
the individual pixels of the two galaxies (see Figure 3). It appears
that the stellar populations in the two objects have different
properties but are relatively homogeneous within each object. The
dispersions of the color distributions  for object E are marginally
larger than the magnitude error in each pixel. On the other hand, the
colors dispersions for object W are about twice as large as the formal
error, confirming that object W consists in a complex mixture of stars
of different ages, dust, and emission line gas.  

In Figure 4 we show a
pixel-by-pixel color-magnitude diagram B vs B-I, presented in the form
of histogram distributions for six magnitude intervals. 
The histograms in the three faintest magnitude intervals were computed
using frames rebinned in larger pixels (3 by 3 of the original ones)
but are still broadened by observational errors (even though only
points with a formal error lower than 0.08 magnitudes were
considered). 
The dotted
line represents the color of the sky background. The thin line
histograms bluer than the sky refer to NGC~454-W, the thin line
histograms redder than the sky represent NGC~454-E. 
Histograms corresponding
to the brighter magnitude intervals are broadened by intrinsic scatter
in color. The color of W varies more than that of E, with the
brightest pixel being bluer; such trend could be due to either age or
reddening effects. 
The thick line
histograms present only in the two fainter intervals correspond to the
area belonging to the west tail of E (area T; see Figure 1). Area T
has an intermediate color between those of the E and W components,
suggesting that it consists in a mixture of stellar populations of
both E and W. The color of the tail is very close to the
color of the sky, even if the former is much brighter than the latter
(the lowest end of the faintest histogram is 0.7 magnitudes brighter
than the sky).  The close similarity in color suggests that a much
more extended and fainter stripped population might occupy vast portions of
the image while being very hard to resolve.

\section{Discussion and Conclusions}

Our observations confirm that the NGC~454 system is in the early
stages of an interaction.  The star-clusters are very young and
contain stars able to provide a very strong UV continuum. It is likely
that the most of the emission lines observed by Johansson (1988) in
object W are actually produced by the star-clusters. The total mass of 
the blue, compact clusters is of the order of 10\% of the whole 
interstellar gas of NGC~454-W, indicating a high efficiency of the 
interaction induced star formation. 

Johansson (1988) emission line spectrum of the E
component, the blue pixels in the central regions, and the
distribution of dust lanes reaching down to the center, provide evidence 
that gas has sunk into the center of component E. However, our observations
do not reveal signs of strong star formation there. Since object E
already follows the correlation between nuclear cusp slope and
magnitude of the spheroidal component observed for other elliptical
galaxies and bulges (Byun et al. 1996), should the two components
merge to form a more massive elliptical galaxy, the cusp slope would
have to decrease for the final merger endproduct to continue obeying
the correlation. Whether the gas in the core of E will ever go through
a star burst phase is hard to predict. Indeed, the emission line
spectrum measured by Johansson with its [OIII]$\lambda 5007$ much
stronger than H$\beta$ and its [NII]$\lambda 6583$ stronger than
H$\alpha$ is more typical of narrow line AGNs than of star forming
regions (Veilleux and Osterbrock 1987).  Thus, it is possible that we
are just witnessing the refueling of a possible AGN engine in the E
component.

\section{Acknowledgements}

We wish to thank Brad Whitmore for a careful reading of the manuscript
and the anonymous referee for useful comments. CMC is supported by
NASA through the grant HF-1079.01-96a awarded by the Space Telescope
Science Institute, which is operated by the Association of
Universities for Research in Astronomy, Inc., for NASA under contract
NAS 5-26555. This research has made use of the NASA/IPAC Extragalactic
Database (NED) which is operated by the Jet Propulsion Laboratory,
Caltech, under contract with NASA.  MR acknowledges support from DDRF
grant 82160.

\newpage
\section{References}

\noindent
Arp H.C., Madore B.F., 1987, ``A Catalogue of Southern Peculiar Galaxies and
Associations'', Cambridge.

\noindent
Barnes J.E., Hernquist L., 1996, ApJ, 471, 115.

\noindent
Brocato E., Castellani V.,1993, ApJ 410, 99.

\noindent
Byun Y.-I., 1996, AJ, 111, 1889.

\noindent
Carollo C.M., Franx M., Illingworth G., Forbes D.A., 1997a, ApJ, 481, 710.

\noindent
Carollo C.M., Danziger I.J., Rich R.M., Chen X., 1997b, ApJ in press.

\noindent
Cassisi S., Castellani V., Straniero O.,1994, AA, 282, 753.

\noindent
Franx M., Illingworth G., Heckman T., 1989, AJ, 98, 538.

\noindent
Holtzman J.A., et al., 1995, PASP, 107, 1065.

\noindent
Iben, I., Jr., 1967, ARAA, 6, 571.

\noindent
Johansson, L., 1988, AA, 191, 29.

\noindent
Kurucz R.L., 1993, ATLAS9 Stellar Atmosphere Programs and 2~$km~s^{-1}$ grid
(Kurucz CD-ROM No. 13)

\noindent
Lauer T.R., et al., 1995, AJ, 110, 2622.

\noindent
Panagia, N., 1973, AJ, 78, 929.

\noindent
Romaniello, M., Panagia, N., Carollo, C. M., Stiavelli, M., 1997, in
preparation.

\noindent
Veilleux S, Osterbrock D.E., 1987, ApJS, 63, 295.

\noindent
Whitmore B.C., Schweizer F., 1995, AJ, 109, 960.

\noindent
Whitmore B.C., Schweizer F., Leitherer C., Borne K., Robert, C., 1993,
AJ, 106, 1354.

\noindent
Worthey G., 1994, ApJS, 95, 107.

\newpage
\section{Figure Captions}

\noindent
{\bf Figure 1:} Left Panel: combined F450W+F606W+F814W image of NGC
454.  The North is identified by the arrow. The boxes identify the areas
discussed in the text. SW, SE, and S are ``point'' sources discussed by
Johansson (1988). We resolve both SW and SE in multiple sources.
Right Panel: F450W-F814W color map.

\noindent
{\bf Figure 2:} NGC~454 East light profiles and colors. On the upper
left panel we show the light profiles in the F450W, F606W, F814W
filters. The bottom panels show the F450W-F606W and the
F606W-F814W color profiles (left and right respectively). 
The upper right panel shows
a B-V vs V-I plot. The dotted line indicates the effect of reddening
while the arrow indicates the effect of line emission in the broad
band filters.

\noindent
{\bf Figure 3:} The F450W-F606W vs F606W-F814W color-color diagram for
the point sources: filled squares, open triangles and crosses denote
points with color errors less than 0.12, 0.12-0.20, and 0.20-0.40,
respectively (63, 77 and 93 sources, respectively). Theoretical
evolutionary tracks (3 Myrs to 5 Gyrs) are displayed as a heavy line.
For reference, the colors of individual pixels of components NGC~454-E
and NGC~454-W are also plotted. The dashed lines represent the effect
of reddening while the dotted line indicates the effect of
contamination by emission lines. The scale of such effects is
identified by the two arrows. The inset shows the luminosity function
in the $V_{F606W}$ magnitude. Histograms have progressively less 
shading with increasing color uncertainty.

\noindent
{\bf Figure 4:} B vs
B-I diagram for each pixel in the selected areas of Figure 1,
for six intervals one magnitude wide. The dotted line
represents the color of the sky. The thin solid line histograms bluer than
the sky refer to the object W, the thin solid line histograms redder
than the sky represent object E, while the thick solid line histogram,
visible only in the two fainter intervals, refers to an area
corresponding to the west tail of E and the north of W. The tail of E has
a color intermediate between those of E and W. 

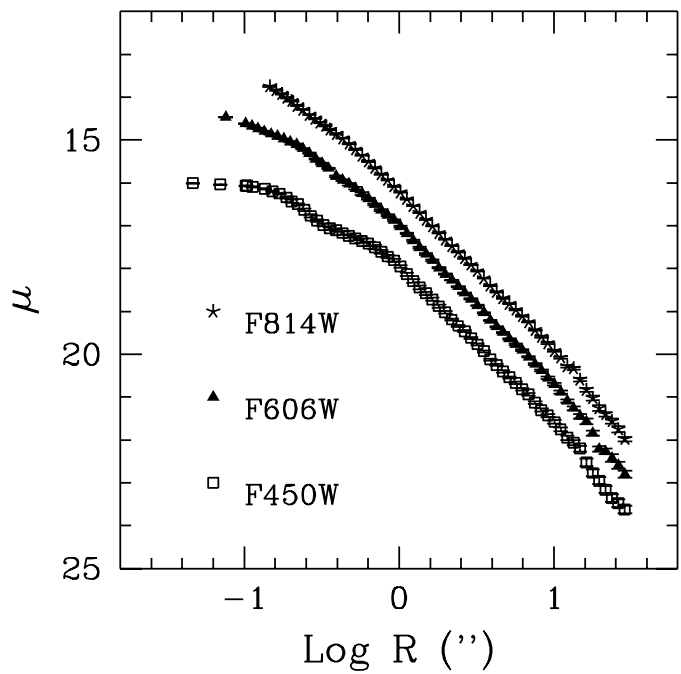
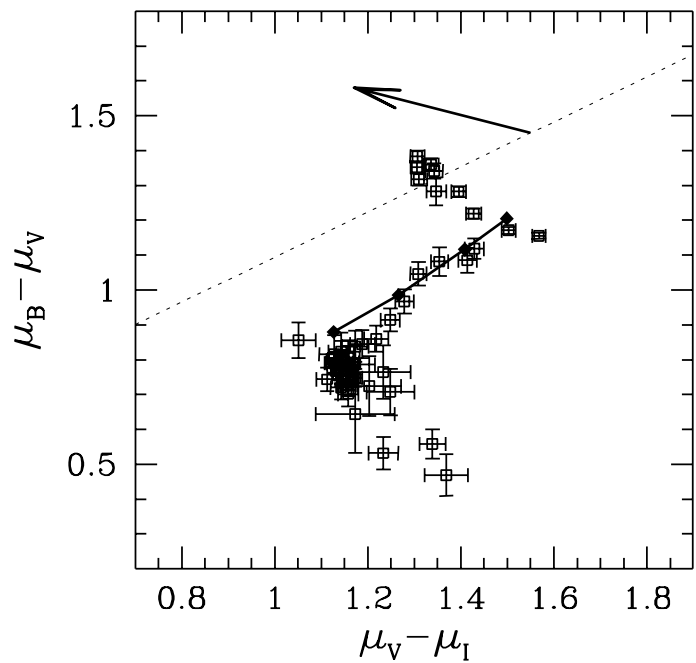
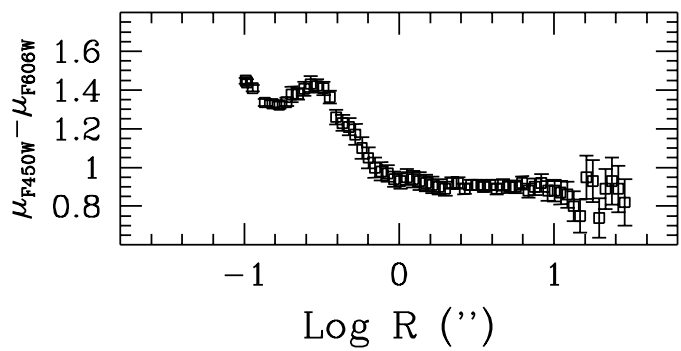
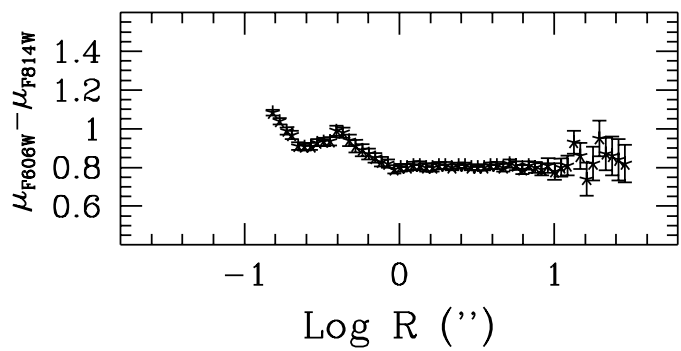
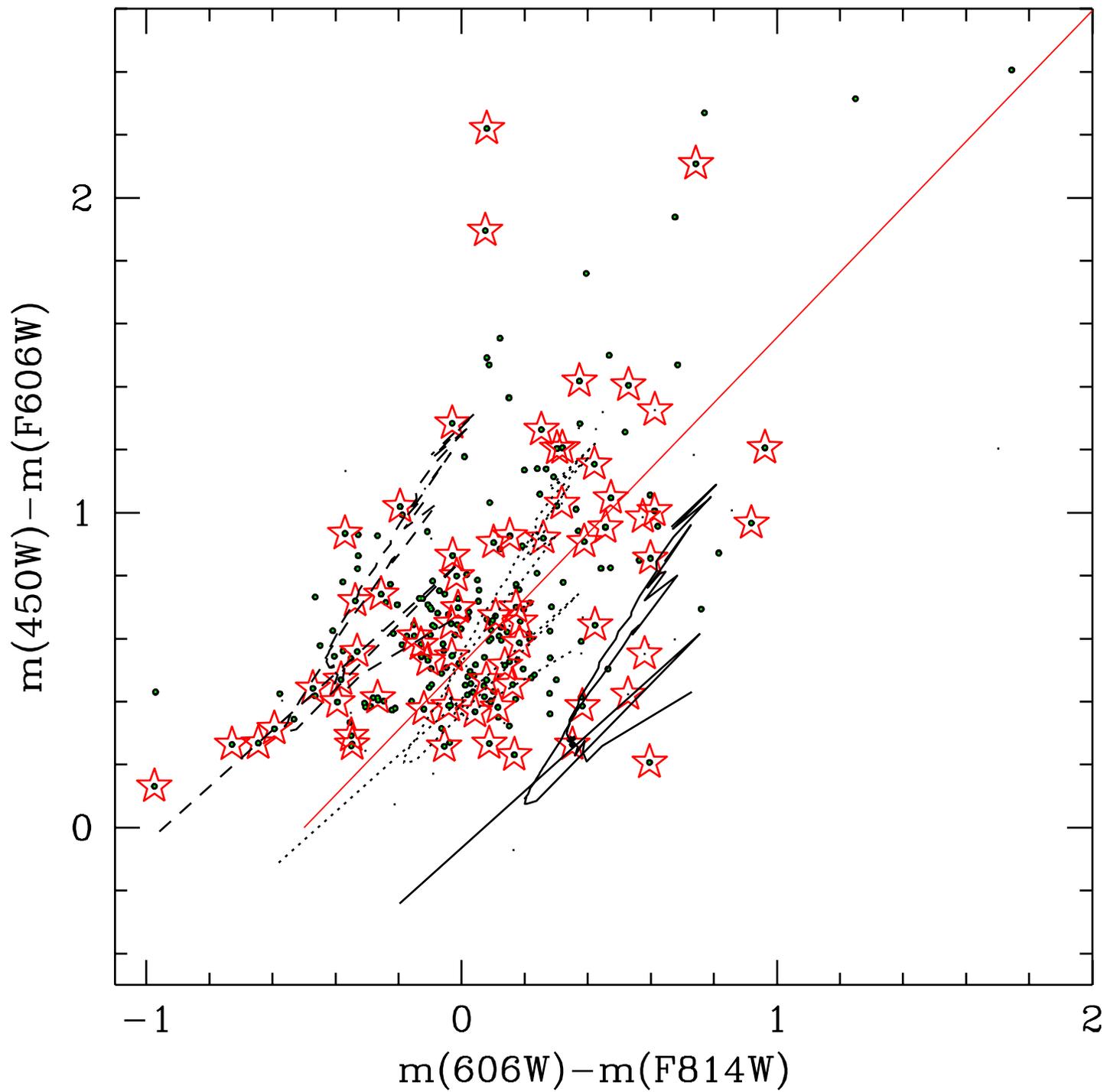
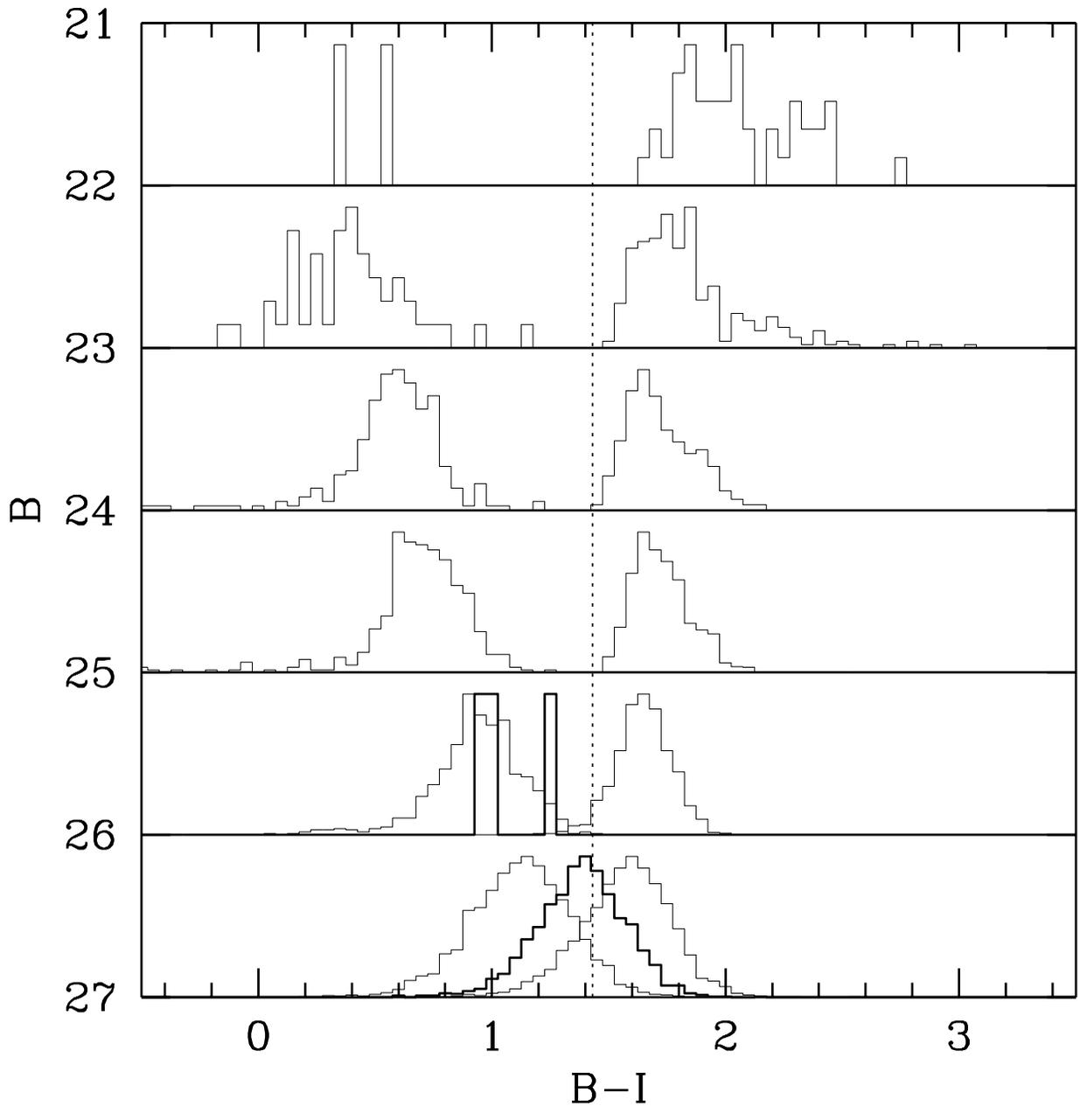
\end{document}